\documentclass[12pt,indent]{article}

\usepackage{a4,latexsym,amsmath,amssymb}
\usepackage[latin1]{inputenc}
\usepackage{float}
\usepackage{natbib}
\setcitestyle{aysep={}} 
\usepackage{graphics}
\usepackage[english]{babel}
\usepackage{tabularx}
\usepackage{lscape}
\usepackage{multirow}
\usepackage{rotating}
\usepackage{dsfont}
\usepackage{caption}
\usepackage{subcaption}
\usepackage{bbm}
\usepackage[table]{xcolor}
\usepackage{booktabs}
\usepackage{calc}
\usepackage{ifthen}
\usepackage{url}
\usepackage{colortbl}      
\usepackage{tikz}

\usepackage{blindtext}
\usepackage{scrextend}
\usepackage{mathtools}
\usepackage{verbatim}

\usepackage{subcaption}
\usepackage{fixltx2e}
\usepackage{booktabs}
\usepackage{threeparttable}
\usepackage{array}
\usepackage{tabulary}
\usepackage{comment}

\usepackage[toc,page]{appendix}
\usepackage{etoolbox}
\newcommand{\nocontentsline}[3]{}
\newcommand{\tocless}[2]{\bgroup\let\addcontentsline=\nocontentsline#1{#2}\egroup}

\usetikzlibrary{shapes.geometric, chains, calc}

\usepackage{natbib}
\usepackage{subcaption}

\usepackage{geometry}
\usepackage{setspace}
\usepackage{tikz}
\usepackage[]{graphicx}
\usepackage[]{color}
\makeatletter
\def\maxwidth{ %
  \ifdim\Gin@nat@width>\linewidth
    \linewidth
  \else
    \Gin@nat@width
  \fi
}
\makeatother

\definecolor{fgcolor}{rgb}{0.345, 0.345, 0.345}

\usepackage{framed}
\makeatletter
 {\par\unskip\endMakeFramed%
 \at@end@of@kframe}
\makeatother

\definecolor{shadecolor}{rgb}{.97, .97, .97}
\definecolor{messagecolor}{rgb}{0, 0, 0}
\definecolor{warningcolor}{rgb}{1, 0, 1}
\definecolor{errorcolor}{rgb}{1, 0, 0}

\usepackage{alltt}
\IfFileExists{upquote.sty}{\usepackage{upquote}}{}

\makeatletter
\renewcommand{\section}{\@startsection{section}{1}{\z@}%
        {-3.5ex \@plus -1ex \@minus -.2ex}%
        {1.5ex \@plus.2ex}%
        {\reset@font\large\sffamily}}
\renewcommand{\subsection}{\@startsection{subsection}{1}{\z@}%
        {-3.25ex \@plus -1ex \@minus -.2ex}%
        {1.1ex \@plus.2ex}%
        {\reset@font\normalsize\sffamily\flushleft}}
\renewcommand{\subsubsection}{\@startsection{subsubsection}{1}{\z@}%
        {-3.25ex \@plus -1ex \@minus -.2ex}%
        {1.1ex \@plus.2ex}%
        {\reset@font\normalsize\sffamily\flushleft}}
\makeatother

\newsavebox{\tempbox}
\newlength{\linelength}
 \makeatletter
\renewcommand{\@makecaption}[2]
{
  \renewcommand{\baselinestretch}{1.1} \normalsize\small
  \vspace{5mm}
  \sbox{\tempbox}{#1: #2}
  \ifthenelse{\lengthtest{\wd\tempbox>\linelength}}
  {\noindent\hspace*{4mm}\parbox{\linewidth-10mm}{\sc#1: \sl#2\par}}
  {\begin{center}\sc#1: \sl#2\par\end{center}}
}

\newcommand{\cov}{\operatorname{cov}}

\newcommand{\alphab}{\boldsymbol{\alpha}}

\newcommand{\betab}{{\boldsymbol{\beta}}}

\newcommand{\deltab}{\boldsymbol{\delta}}

\newcommand{\gammab}{\boldsymbol{\gamma}}

\newcommand{\Sigmab}{\boldsymbol{\mathit\Sigma}}

\newcommand{\vb}{\boldsymbol{v}}

\newcommand{\wb}{\boldsymbol{w}}

\newcommand{\xb}{\boldsymbol{x}}

\newcommand{\yb}{\boldsymbol{y}}

\newcommand{\zb}{\boldsymbol{z}}

\newcommand{\blanco}[1]{}

\newcommand{\defas}{\mathrel{\mathop{:}}=}
\renewcommand{\sb}{\boldsymbol{s}}

\begin{document}
\sloppy
\blanco{
\thispagestyle{empty}
\begin{center}
\vspace{0.3cm}
\Large{Heterogeneity in General Multinomial Choice Models}
\end{center}

\vspace{1.5cm}
\begin{center}
\large{Ingrid Mauerer} \\
\vspace{0.2cm}
\small{
University of Barcelona \\
School of Economics \\
Institutions and Political Economy Research Group \\
Address: John M. Keynes, 1-11, 08034 Barcelona, Spain \\
Phone: +34 93 403 72 32  \\
Email: ingridmauerer@ub.edu \\
 }

\vspace{1cm}
\large{Gerhard Tutz} \\
\vspace{0.2cm}
\small{
LMU Munich \\
Department of Statistics \\
Address: Akademiestra{\ss}e 1, 80799 M\"{u}nchen, Germany\\
Phone: +49 89 2180 3044\\
Email: tutz@stat.uni-muenchen.de  \\
 }
\end{center}
\vspace{3cm}
}
\title{Heterogeneity in General Multinomial Choice Models}
%
\author{\large{Gerhard Tutz} \\
\small{LMU Munich, Department of Statistics} \\
\small{Address: Akademiestra{\ss}e 1, 80799 M\"{u}nchen, Germany}\\
\small{Email: tutz@stat.uni-muenchen.de }
\vspace{0.8cm}\\
\large{Ingrid Mauerer} \\
\small{University of Barcelona, School of Economics} \\
\small{Institutions and Political Economy Research Group} \\
\small{Address: John M. Keynes, 1-11, 08034 Barcelona, Spain} \\
\small{Email: ingridmauerer@ub.edu} 
}
%

\date{ \today}
\maketitle
\thispagestyle{empty}
\onehalfspacing

\begin{abstract} 
\noindent
Different voters behave differently, different governments make different decisions, or different organizations are ruled differently. Many research questions important to political scientists concern choice behavior, which involves dealing with nominal-scale dependent variables.  Drawing on the principle of maximum random utility, we propose a  flexible and general heterogeneous multinomial logit model for studying differences in choice behavior. The model systematically accounts for heterogeneity that is not captured by classical models, indicates the strength of heterogeneity, and permits examining which explanatory variables cause heterogeneity. As the proposed approach allows incorporating theoretical expectations about heterogeneity into the analysis of nominal dependent variables, it can be applied to a wide range of research problems. Our empirical example uses data on multiparty elections to demonstrate the benefits of the model in the study of heterogeneity in spatial voting.
\end{abstract}


\thispagestyle{empty}
\clearpage
\pagenumbering{arabic}
\setcounter{page}{1}

\tocless
\section{Introduction}
\noindent
Many research questions in political science are categorical in nature.  Regression models for categorical dependent variables are well-established and widely applied in the discipline to analyze research problems that involve two or more categories without an ordering structure \citep[see, e.g., ][]{Agresti.2007, Long.1997,tutzbuch12}.   Statistical techniques belonging to this model class build an established methodological subfield in the discipline. Various aspects, features, and key methodological problems that arise when dealing with categorical dependent variables have been discussed to enhance and simplify their applications.  The methodological contributions comprise several approaches to measure and visualize the goodness of fit \citep[e.g.,][]{Esarey.2012, Greenhill.2011, Hagle.1992, Herron.1999},  address the separation problem \citep{ Cook.2018, Rainey.2016, Zorn.2005}, or discuss the evaluation of interactive hypotheses \citep{Berry.2010} in such models.

Although many research questions in political science involve theoretical expectations about heterogeneous effects, there are little efforts in allowing heterogeneity in categorical dependent variables.  The most prominent way to relax the homogeneity assumption for nominal-scaled dependent variables is the mixed logit model (MXL) \citep[see, e.g.,][]{Greene.2006, McFadden.2000b}, which is applied to study heterogeneity in government choice \citep{Glasgow.2012, Glasgow.2015} or voting behavior \citep{Glasgow.2001}.  However, for researchers, the MXL model can be quite demanding to apply. For example, the researcher needs to decide on a distribution for the subject-specific heterogeneity to approximate the underlying behavioral process, and repeated measurements are necessary to identify the model.

In this paper, we propose a methodological approach that is very flexible and general in accounting for heterogeneity in nominal-scale dependent variables. Relying on the random utility maximization framework, we derive a multinomial logit model, called the \textit{General Heterogeneous Multinomial Logit Model} (GHMNL),  which allows for systematically studying heterogeneity in choice behavior. The proposed model builds on the standard multinomial logit model (MNL), also known as the conditional logit model \citep[][]{McFadden.1974, Yellott.1977}, which is the most frequently applied statistical tool to study choices among discrete alternatives.\footnote{Throughout the paper, we use the term MNL to refer to multinomial logit models that contain covariates that depend on the outcome categories and those that do not.} 

As the MNL model, the GHMNL model is a classical discrete choice model that can handle both choice-specific and chooser-specific explanatory variables. In contrast to the MNL model, which ignores that the variance of the underlying latent traits can be chooser-specific, the GHMNL model accounts for such heterogeneous effects.  The extension integrates a heterogeneity term into the systematic part of the utility function.  The heterogeneity term is linked to explanatory variables and permits accounting for behavioral tendencies in choice behavior without referring to latent variables. It provides an indicator of the degree of distinctiveness of choice, indicates the strength of heterogeneity in choice behavior, and allows examining which explanatory variables cause heterogeneity. Therefore, the proposed model enables incorporating theoretical expectations about heterogeneity into the analysis of nominal dependent variables.  As compared to the MXL model, the GHMNL model also comes with convenient properties and assumptions, such as its closed-form solution for evaluating the outcome probabilities. In addition, the GHMNL model frees the researcher from making distributional assumptions for the random parameters and is computationally straightforward. 

We apply the model to electoral choices in multiparty elections and demonstrate its benefits in the study of heterogeneity in spatial voting. This empirical application has several merits. First, spatial voting models typically contain both types of explanatory variables, choice-specific (voter-party issue proximities) and chooser-specific (socioeconomic voter attributes) ones. Second, the literature on voter heterogeneity provides several theoretical concepts why not all voters assign the same importance to issue considerations, including, for instance, platform divergence or political sophistication \citep[e.g.,][]{Campbell.1960, Luskin.1987, RePass.1971}.  We will demonstrate how the proposed model allows incorporating such theoretical expectations into the empirical modeling. Although we focus on electoral choices and voter heterogeneity in our empirical application, we see great potential for applying the model to explore heterogeneous effects in all sub-disciplines, such as in the study of legislative behavior, public opinion and attitudes, international relations, or comparative politics.  

Based on a brief review of the classical discrete choice model, we first derive our general heterogeneous multinomial choice model and outline how it extends the standard MNL model. Next, we investigate the differences between the general heterogeneous multinomial choice model and the MXL model. Then, we demonstrate the usefulness of our model by examining heterogeneity in spatial issue voting.

\vspace{1cm}
\tocless
\section{The Standard Multinomial Choice Model}
\noindent
The multinomial logit model (MNL) is the most common model to study choice behavior \citep[see, e.g.,][]{ Hensher.2015, Louviere.2009, Train.2009}. One key feature of the MNL model limits our insights into heterogeneity in choice behavior. It ignores that the variances of the underlying latent traits can vary across decision makers.   A brief review of the MNL model will help to motivate the model we propose and its advantages. 

In the following,  $Y_i \in \{1, \dots, J\}$ will denote the dependent variable that consists of $J$ unordered multiple categories for $i \in \{1,\ldots, n\}$ observations. Within the discrete choice framework, the categories represent $J$ discrete, mutually exclusive, and finite alternatives of which decision makers  choose one. The choice outcome can be a function of  two types of explanatory variables: \textit{choice-specific} and \textit{chooser-specific} variables. The former are variables that are specific for each category and therefore take different values across both alternatives and choosers. They characterize the choice alternatives,  such as price or distance in a classical mode choice situation. Let the choice-specific variables be denoted by $z_{ijk}$, $j \in \{1, \dots, J\}$, $k\in\{1,\ldots,K\}$.  Chooser-specific variables contain characteristics of the decision makers, which vary over decision makers but are constant across the alternatives, such as age or gender.  Let $s_{im}, m\in\{1,\ldots,M\}$ denote the chooser-specific covariates.   

A common way to motivate a choice model is to consider the utilities associated with the alternatives as latent variables.  Let $U_{ij}$ denote an unobservable random utility that  represents how attractive or appealing each  alternative  $j \in \{1, \ldots, J\}$ is for chooser $i \in \{1,\ldots, n\}$. The decision makers are assumed to assess and compare each alternative and select the one that maximizes the random utility so that  $Y_i$   is linked to the latent variables by the \textit{principle of maximum random utility}, 
\[
Y_i=j\quad\Leftrightarrow\quad U_{ij}=\max_{s\in \{1,\dots,J\}} U_{is}.
\]
In a random utility framework, the utility is determined by $U_{ij} = V_{ij} + \varepsilon_{ij}$, where  $V_{ij}$  represents the systematic part of the utility, specified by explanatory variables and unknown parameters, whereas $\varepsilon_{i1},\dots, \varepsilon_{iJ}$ are independent and identically distributed (i.i.d.) random variables with distribution function $F(.)$.

The systematic part of the utility function is specified as a linear predictor
\begin{equation}\label{eq:V_MNL}
V_{ij} = \beta_{j0} +  \sum_{k=1}^K  z_{ijk}\alpha_{k} + \sum_{m=1}^M s_{im}\beta_{jm} = \beta_{j0} + \zb_{ij}^T \alphab + \sb_{i}^T \betab_j ,
\end{equation}
where
\begin{itemize}
\item[ ] 
 $\beta_{10},\ldots,\beta_{J0}$ are the alternative-specific constants.
\item[ ] 
 $\alphab^T=(\alpha_1, \dots, \alpha_K)$ are the parameters\footnote{For simplicity, we assume that the parameters $\alphab$ are identical for all alternatives, i.e., $\alphab_1 = \ldots = \alphab_j \defas \alphab$. This simplification results in a so-called generic or global effect, which does not depend on the alternatives.  See  \cite{Mauerer.2015, Mauerer.2016, Thurner.2000b} for relaxation of the assumption in the study of spatial voting in multiparty elections, and see \cite{Mauerer.2015b} for a parameter selection procedure to systematically reduce the resulting model complexity.}  associated with the vector of choice-specific variables $\zb_{ij}^T=(\zb_{ij1},\dots,\zb_{ijK})$, which indicate the weight decision makers attach to each  attribute $k$ of the alternatives.
\item[ ]  
 $\betab_j^T=(\beta_{j1}, \dots, \beta_{jM})$ is a coefficient vector that  expresses how the  chooser attributes contained in $\sb_{i}^T=(s_{i1} ,\dots,s_{iM})$ determine the choice. 
\end{itemize}

By assuming  that $\varepsilon_{i1},\dots,\varepsilon_{iJ}$  are i.i.d. variables with distribution function $F(x) = \exp(-\exp(-x))$, which is known as the Gumbel  or maximum extreme value distribution, one obtains the classical standard multinomial logit model \citep[see][]{McFadden.1974, Yellott.1977} 
\begin{equation}\label{eq:Prob_MNL}
P(Y_i=j| \{\zb_{ij}\}, \sb_i)= \frac{\exp(V_{ij})}{\sum_{s=1}^J \exp(V_{is})}=  \frac{\exp(\beta_{j0} + \zb_{ij}^T \alphab + \sb_{i}^T \betab_j)}{\sum_{s=1}^J \exp(\beta_{s0} + \zb_{is}^T \alphab + \sb_{i}^T \betab_s)}, 
\end{equation}
$j \in \{1, \dots, J\}$. Since the chooser-specific variables $\sb_i$ are constant over the alternatives, not all of the corresponding coefficients are identifiable. The same applies to the constants. To identify the model,   side constraints are needed. We will use the standard side constraint based on a reference alternative, whose coefficients are set to zero. We select the first alternative as reference and set $\beta_{j0}=0$ and $\betab_1^T=(0,\dots,0)$. 

The standard MNL model presented in Equations (\ref{eq:V_MNL}) and (\ref{eq:Prob_MNL})  ignores that the variance of the underlying latent traits can be subject-specific so that the variances are not allowed to differ across decision makers. Previous research has shown that ignoring variance heterogeneity can yield biased estimates \citep[see, e.g.,][]{TuHetchoice}. 

\tocless
\section{A General Heterogeneous Multinomial Choice Model}
\noindent
In this section, we derive a general multinomial choice model, called \textit{General Heterogeneous Multinomial Logit Model}, in short GHMNL, that accounts for variance heterogeneity in choice behavior. The GHMNL model builds on the model in \citet{TuHetchoice}, which is restricted to global covariates that do not depend on the outcome categories. By contrast, the approach we propose explicitly incorporates choice-specific explanatory variables, which lay at the heard of discrete choice models as attributes of the choice alternatives are the source of utility in discrete choice models. In addition, we outline in detail the interpretation of the novel heterogeneity term we incorporate into the utility function and the estimation methods. In the following, we begin by describing the specification of the utility functions and the choice probabilities in the GHMNL model. 

\vspace{1cm}
\tocless
\subsection{Utility Functions and Choice Probabilities}
\noindent
The GHMNL model extends the standard MNL model by adding a heterogeneity term to the systematic part of the utility function.  For simplicity, let all the  explanatory variables and the constants be collected in  the alternative-specific vector  $\xb_{ij}^T=(\boldsymbol{1}_j^T,\boldsymbol{0}, \dots,\zb_{ij}^T,\dots, \boldsymbol{0})$, where $\boldsymbol{1}_j$ is the $j$th unit vector and $\boldsymbol{0}$ is a vector of zeros. Then, the utility functions take the form 
\[
V_{ij} =  \beta_{j0} + \zb_{ij}^T \alphab + \sb_{i}^T \betab_j = \xb_{ij}^T \deltab,
\]
where $\deltab^T=(\beta_{10},\dots,\beta_{J0},\alphab^T,\betab_1^T,\dots,\betab_J^T)$.  To derive the GHMNL model, we assume  that the latent utilities are given more generally by 
\[
 U_{ij}=\xb_{ij}^T\deltab+\sigma_i\varepsilon_j, 
\]
where $\sigma_i$ is the standard deviation associated with decision maker $i$.  

In the GHMNL model, the standard deviation is linked to explanatory variables  by assuming $\sigma_i= e^{-\wb_i^T\gammab}$,  where $\wb_i$ is a vector of chooser-specific covariates and $\gammab$ is a vector of parameters. As a result, the utility $V_{ij}$  is specified as 
\begin{equation}\label{eq:V_GHMNL}
V_{ij}=\xb_{ij}^T\deltab e^{\wb_i^T\gammab}= (\beta_{j0}+ \zb_{ij}^T\alphab+\sb_{i}^T\betab_j )e^{\wb_i^T\gammab},
\end{equation} 
where 
\begin{itemize}
\item[ ]
$\sb_{i}$ is a vector of chooser-specific covariates, and $\zb_{ij}$ is a vector of alternative-specific covariates. As in the standard MNL model, the variables $\sb_{i}$ have alternative-specific effects and $\zb_{ij}$ global effects. 
\item[]
$\wb_i^T=(w_{i1},\dots,w_{iL})$ is a vector of chooser-specific variables, which can be a subset of $\sb_{i}$. It contains attributes of the decision makers that are supposed to cause heterogeneity in choice behavior. The corresponding parameter vector $\gammab^T=(\gamma_{1}, \dots, \gamma_{L})$ indicates the strength  of  heterogeneity in choosing one alternative. 
\end{itemize}
The model distinguishes between two types of effects: a \textit{location effect} and a \textit{heterogeneity effect}. The term $\xb_{ij}^T\betab_j$ in Equation (\ref{eq:V_GHMNL}) represents the location effect. It is also present in the standard MNL model and determines which alternative the chooser tends to prefer. The novel term  $\wb_i^T\gammab$ represents the heterogeneity effect that determines the impact of heterogeneity in choice behavior.
 
As  the standard MNL model, the GHMNL model has  a closed-form solution for evaluating the choice probabilities so that the utility functions $V_{ij}$ are linked to the choice probabilities through a logistic response function, 
\begin{equation}\label{eq:GHMNL}
P(Y_i=j|\{\xb_{ij}\},\wb_i)=\frac{\exp(\xb_{ij}^T\deltab e^{\wb_i^T\gammab})}{\sum^J_{s=1}\exp(\xb_{is}^T\deltab e^{\wb_i^T\gammab})}, \: j \in \{1, \dots, J\}.
\end{equation} 
Alternatively, the relationship between the choice probabilities and the utility functions can be expressed in terms of 
odds: 
\begin{equation*}
\begin{aligned}
  \frac{P(Y_i=j|\{\xb_{ij}\},\wb_i)}{P(Y_i=s|\{\xb_{ij}\}, \wb_i)} =& \exp\{(\xb_{ij}- \xb_{is})^T \deltab e^{\wb_i^T\gammab}\} \\
	=& \exp\{[\beta_{j0}-\beta_{s0}+(\zb_{ij}- \zb_{is})^T\alphab+\sb_i(\betab_j - \betab_s)]e^{\wb_i^T\gammab}\}. 
	\end{aligned}
\end{equation*}

\vspace{1cm}
\tocless
\subsection{Interpretation of the Heterogeneity Term}
\noindent
The essential novel term in the GHMNL model is the heterogeneity term. It is modeled by the factor $e^{\wb_i^T\gammab}$ and represents the (inverse) standard deviation of the latent variables. The heterogeneity term can be understood as representing variance heterogeneity. However, it also allows for an interpretation without reference to latent variables, which are always elements used to build a model but cannot be observed.  The heterogeneity term represents a specific choice behavior that permits accounting for behavioral tendencies that are not linked to particular alternatives:
\begin{itemize}
\item[ ]
When $\wb_i^T\gammab\rightarrow -\infty$, one obtains $P(Y_i=j|\{\xb_{ij}\},\wb_i)=1/J$. In this extreme case,  all alternatives have the same choice probabilities. It implies that the decision maker chooses an alternative at random because none of the covariates can systematically explain the choice.  The chooser shows maximal heterogeneity.
\item[ ]
When $\wb_i^T\gammab\rightarrow \infty$ and  the condition $\xb_{ij}^T\betab_j \ne 0$ holds at least for one $j > 1$,  the probability for one of the $j \in \{1,\dots,J\}$ alternatives approaches 1. In this case,  the decision maker has a distinct preference, and shows minimal heterogeneity. Therefore, choosers with large $\wb_i^T\gammab$-values  show less variability, they distinctly prefer specific alternatives. 
\end{itemize}
Thus, the heterogeneity term $\wb_i^T\gammab$ can be considered as an indicator of the degree of distinctness of choice or as a measure of heterogeneity in choice behavior.  For small values of $\wb_i^T\gammab$, the difference between the choice probabilities becomes small. By contrast, the difference between a specific alternative and the remaining ones gets larger when  $\wb_i^T\gammab$ increases.  As the heterogeneity term contains attributes of the decision makers, the model systematically accounts for heterogeneity in choice behavior across individuals. It allows examining which explanatory variables cause heterogeneous effects. For example, suppose $w_{i}$ denotes age and $\gamma$ is positive. It would suggest that older decision makers have more clear cut preferences than younger ones. The former tend to prefer specific alternatives, while younger decision makers have less distinct preferences and show more heterogeneity in selecting one alternative.

\begin{figure}[h!]
\caption{Illustration of the Heterogeneity Term in the GHMNL Model}
\label{fig:1}
\begin{center}
\begin{subfigure}[c]{0.8\textwidth}
\subcaption{Positive $\gamma_1$-values:  $\square $  $\gamma_1=1$,  $\triangle $ $\gamma_1=3$} 
\includegraphics[width=1\textwidth]{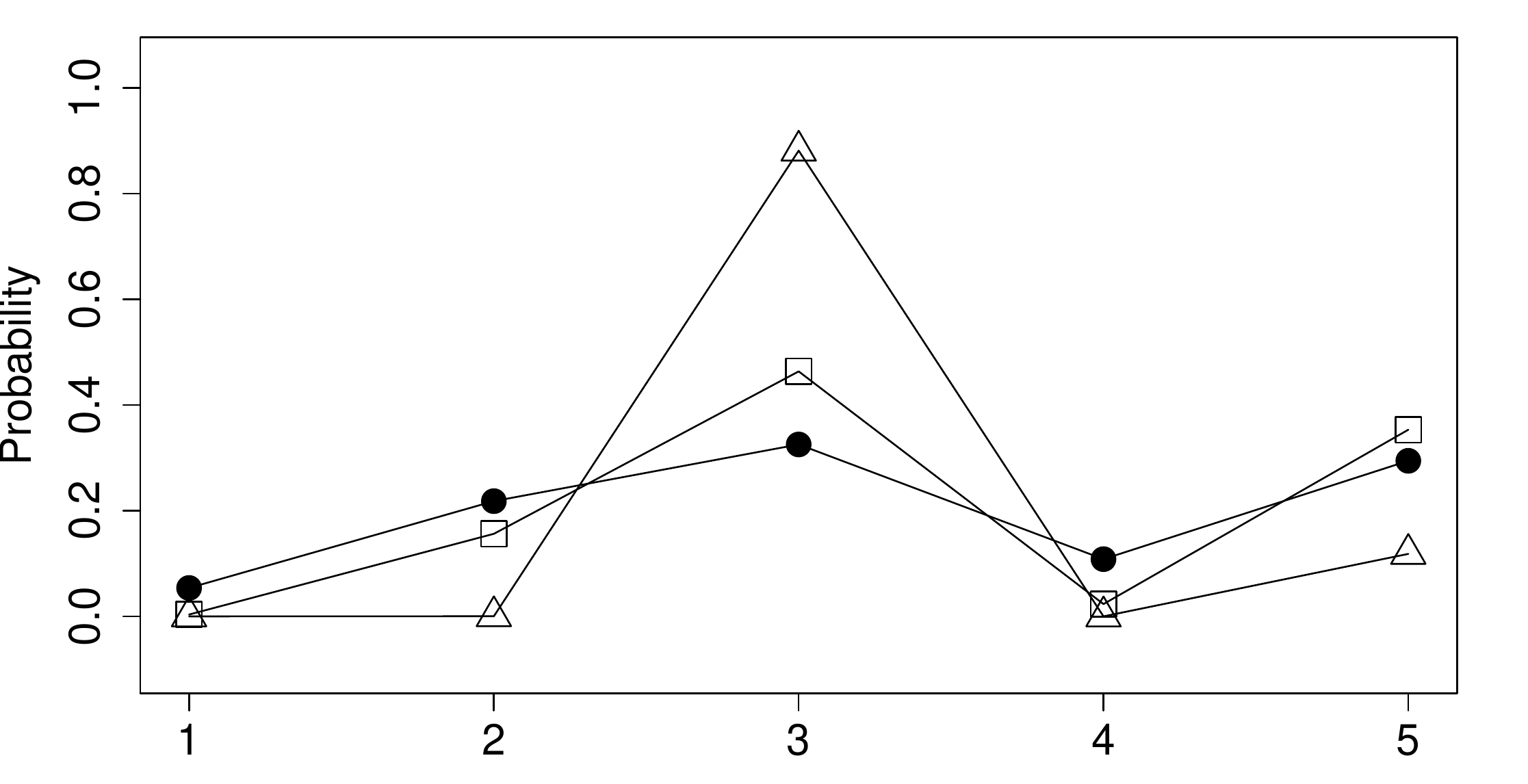} 
\end{subfigure}
\vspace{0.5cm}
\begin{subfigure}[c]{0.8\textwidth}
\vspace{0.5cm}
\subcaption{Negative $\gamma_1$-values:  $\square $ $\gamma_1=-1$,  $\triangle $ $\gamma_1=-3$}
\includegraphics[width=1\textwidth]{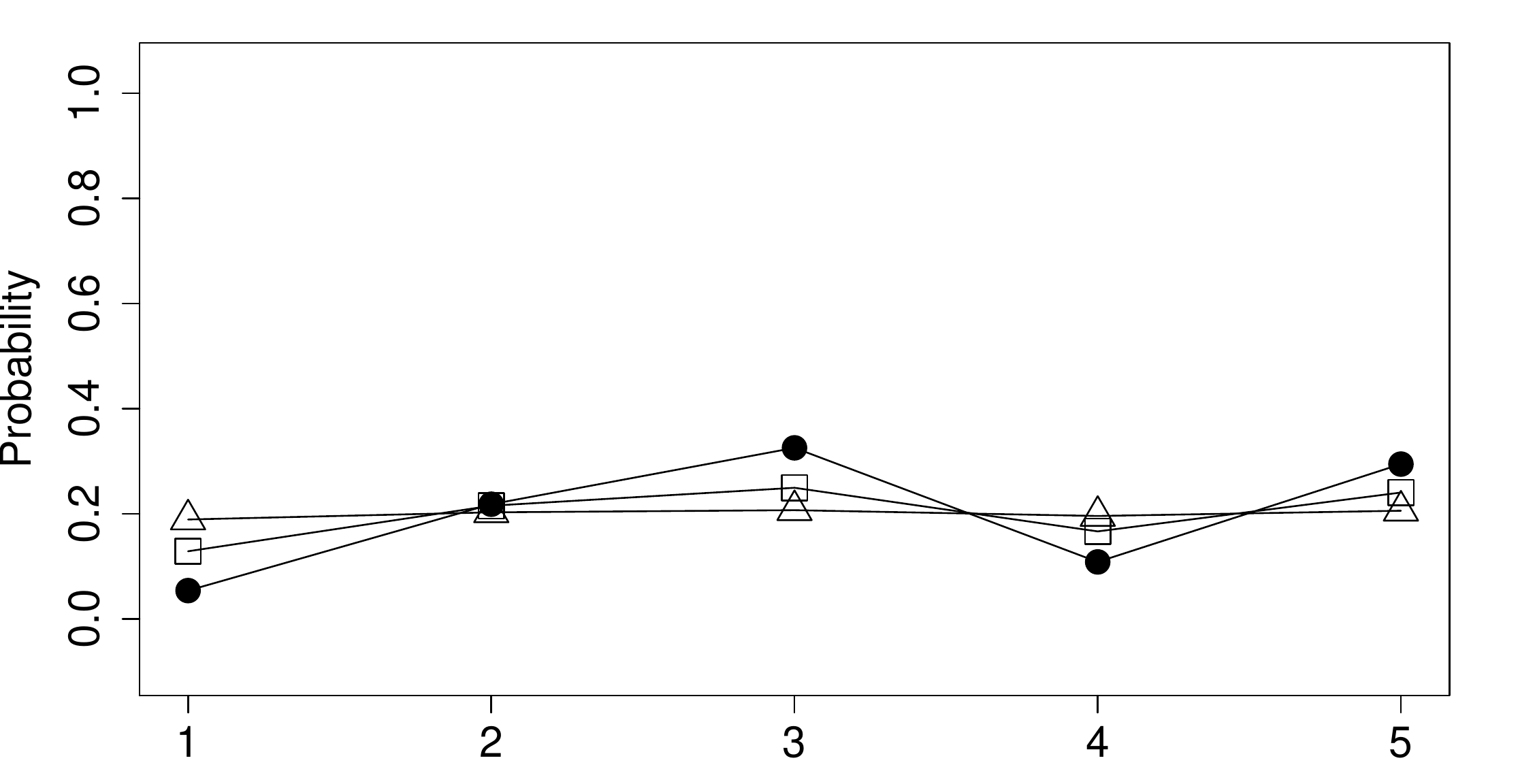} 
\end{subfigure}
\end{center}
{\small \textit{Note:} Figures depict the probabilities $P(Y_{i}=j)$ for a model with one binary and one quantitative normally distributed covariate for five alternatives  $j \in \{1, 2,3,4,5\}$. Panel (a) shows the effects for positive $\gamma_1$-values, panel (b) for negative $\gamma_1$-values. In both panels, the filled circles $\bullet$ depict the base probabilities when no heterogeneity is present ($\gamma_1=0$). The probability curves result by plugging the respective values into Equation (\ref{eq:GHMNL}). }
\end{figure}

Figure \ref{fig:1} illustrates the behavioral tendencies the GHMNL model can uncover. For a five-choice situation $j \in \{1, 2,3,4,5\}$, it depicts the probabilities $P(Y_{i}=j)$ for a model with two covariates contained in the heterogeneity term $\wb_i$, one binary and one quantitative normally distributed explanatory variable.  For the binary covariate, we consider the effect at value $\wb_i^T =(1,0)$.  The two panels show the probabilities for different parameter values ($\gamma_1$) in the heterogeneity term: panel (a) shows the effects for positive $\gamma_1$-values, panel (b) for negative $\gamma_1$-values. In both panels, the filled circles depict the probabilities that result when no heterogeneity is present, that is, when $\gamma_1=0$, resulting in the standard MNL model.  

When inspecting the base probabilities obtained from the standard MNL model, we see that the decision maker prefers alternative 3, and to a lesser extent alternative 5.  Panel (a) shows that this pattern becomes more pronounced for increasing $\gamma_1$-values. Thus, the decision maker more distinctly prefers alternative 3 in the GHMNL model.  By contrast, the pattern flattens for negative $\gamma_1$-values, as illustrated in panel (b).  This indicates that the decision maker tends to choose an alternative at random and shows substantial heterogeneity in selecting one of the five alternatives. 

\tocless
\subsection{Estimation}
\noindent
In the following, we outline how the parameters  of the GHMNL model can be estimated. Let the choice outcome be represented as a vector $\yb_i = (y_{i1}, \dots, y_{iJ})^T$ with $y_{i1}$ taking the value  1 when alternative $j \in \{1, \dots, J\}$ is chosen and 0 otherwise so that one obtains $\yb_i = (0,\dots,0,1,0,\dots,0)^T$  if  $Y_i = j$.  Let  $\pi_{ij}=P(Y_i=j|\{\xb_{ij}\},\wb_i)$ denote the choice probabilities and  $\deltab^T=(\beta_{10},\dots,\beta_{J0},\alphab^T,\betab_1^T,\dots,\betab_J^T,\gammab^T)$ the overall parameter vector that collects all coefficients to be estimated.

Using the first alternative as reference, the kernel of the log-likelihood of the model presented in Equation (\ref{eq:GHMNL}) is given by 
\begin{equation*}
\begin{aligned}
l(\deltab)= &  \sum_{i=1}^{n}\{ \sum_{j=2}^{J} y_{ij}
\log \left(\frac{\pi_{ij}} {1-\pi_{i2}-\dots-\pi_{iJ}}\right) +
\log(1-\pi_{i2}-\dots-\pi_{iJ})\} \\
=& \sum_{i=1}^{n}\{ \sum_{j=2}^{J} y_{ij}
(\xb_{ij}^T\deltab) e^{\wb_i^T\gammab} -
\log(1+ \sum_{s=2}^{J}\exp(\xb_{is}^T\deltab e^{\wb_i^T\gammab})\} .
\end{aligned}
\end{equation*}

For the maximization of the log-likelihood, we make use of  the first derivatives, also known as score functions. They take the form 
\begin{align*}
\partial l(\deltab) / \partial \delta_t = 
&\sum_{i=1}^{n}\{ \sum_{j=2}^{J} y_{ij}
x_{ijt}\delta_t e^{\wb_i^T\gammab} 
-\frac{\sum_{s=2}^{J} x_{ist} e^{\zb_i^T\gammab} \exp(\xb_{is}^T\deltab e^{\wb_i^T\gammab})} {1+ \sum_{s=2}^{J}\exp(\xb_{is}^T\deltab e^{\wb_i^T\gammab})}\},
\\
\partial l(\deltab) / \partial \gamma_t = 
&\sum_{i=1}^{n}\{ \sum_{j=2}^{J} y_{ij}
(\xb_{ij}^T\deltab) e^{\wb_i^T\gammab} w_{it} 
-\frac{\sum_{s=2}^{J} \xb_{is}^T\deltab e^{\wb_i^T\gammab} z_{it}\exp(\xb_{is}^T\deltab e^{\wb_i^T\gammab})} {1+ \sum_{s=2}^{J}\exp(\xb_{is}^T\deltab e^{\wb_i^T\gammab})}\}.
\end{align*}
As approximation of the covariance $\cov(\hat\deltab)$, we use the observed information $-\partial^2 l(\hat\deltab)/\partial \deltab \partial \deltab^T$.

\blanco{
Ridge penalty (double)
\begin{align*}
&J(\deltab,\gammab) = \lambda_1\sum_j (Ind_j \delta_j)^2 + \lambda_2\sum_j \gamma_j^2,\\
&\partial J(\deltab,\gammab)/\partial \delta_j = \lambda_1 2Ind_j \delta_j,\\
&\partial J(\deltab,\gammab)/\partial \delta_j = \lambda_2 2\gamma_j. 
\end{align*}
where $Ind_j$ is an indicator function with $Ind_j=1$ if $\delta_j$ refers to a variable, and $Ind_j=0$ if $\delta_j$ refers to an intercept.
}

\vspace{1cm}
\tocless
\subsection{Implementation in R}
\noindent
We have written an R function that allows the user  to fit the GHMNL model. Section \ref{app:A} in the Supporting Information describes the 
routines to implement the model. 

\vspace{1cm}
\tocless
\section{The GHMNL Model Contrasted with the Mixed Logit Model}
\noindent
A model that has been used to study heterogeneity in decision behavior is the mixed logit model (MXL) \citep[see][]{Greene.2006, Hensher.2003, McFadden.2000b}.\footnote{The MXL is also referred to as random parameters logit, mixed multinomial logit, or hybrid logit model. We use the most popular term mixed logit model.} The MXL model has been applied in transportation economics and econometrics, and also political science settled on the model to examine heterogeneity in government choice \citep{Glasgow.2012, Glasgow.2015} or voting behavior \citep{Glasgow.2001}. A brief review and discussion of the MXL model illustrate the limitations of this approach and the advantages of the model we propose to account for heterogeneity in choice behavior.  

\vspace{1cm}
\tocless
\subsection{Mixed Logit Model Formulation}
\noindent
Following \citet{Greene.2006},  the MXL model can be derived from latent utilities 
\[
U_{ijt} = \zb_{ijt}^T \alphab_i + \varepsilon_{ijt},
\]
where 
\begin{itemize}
\item[ ]
the additional index $t$ refers to the choice situation, and
\item[ ]
$\zb_{ijt}$ is the full vector of explanatory variables, including attributes of the alternatives, socioeconomic characteristics of the decision makers, and the choice task itself.
\end{itemize}
As compared to the standard MNL model, the crucial extension in the MXL model is that the parameter vector $\alphab_i$ is subject-specific so that the effects are allowed to vary across decision makers $i$.  By assuming that the subject-specific effects are random and in part determined by an additional vector of covariates $\wb_i$, the model becomes a mixed-effects model. The subject-specific effects are assumed to take the form
\[
\alphab_i = \alphab +\Delta \wb_i + \boldmath{\Sigma}^{1/2}\vb_i,
\]                                                              
where 
\begin{itemize}
\item[ ]
$\Delta$ is a matrix of coefficients associated with the covariate vector $\wb_i$, 
\item[ ]
$\vb_i$ is a random  vector of uncorrelated random variables with known variances, 
\item[ ]
$\boldmath{\Sigma}^{1/2}$ is a covariance matrix that determines the variance structure of the random term.
\end{itemize}         
Maximum simulated likelihood estimates are obtained by maximizing the log-likelihood with respect to all the unknown parameters \citep[see also][]{Train.2009}.

By allowing parameters to vary randomly over decision makers instead of assuming that they are the same for every chooser, the MXL model is very flexible and can account for a rather general form of heterogeneity. However, this flexibility comes with the cost of a large number of parameters, which might render estimates unstable without careful variable selection. Further drawbacks of the model are that one has to specify a specific distribution for the subject-specific random effects and the model parameters may not be identified without repeated measurements, that is, without having varying choice situations for the same chooser.

\tocless
\subsection{Comparing Modeling Approaches}
\noindent
Both the GHMNL model and the MXL model can be derived from latent utilities. The main difference between both approaches lies in the motivation of heterogeneity in choice behavior.  In the proposed  GHMNL model, the variances of the latent utilities are allowed to vary across decision makers. By contrast, the MXL model permits the parameters to vary across individual choosers; however,  without further motivation. While the GHMNL model also allows parameters to vary across choosers, it does so in a more restrictive and systematic way.  Here, the effect parameters associated with the alternative-specific covariates are $\alphab e^{\wb_i^T\gammab}$.  Under this specification, the covariates contained in $\wb_i$ modify the effects. Depending on the value of $\wb_i$, the effect is strengthened or weakened. In addition, the same effect modification applies to all coefficients, which is a consequence of the derivation from the variances of the latent utilities. By contrast, the MXL model allows for all sorts of parameter variation, including random variation and even a possible reversal of the sign of effects.   

By allowing the effects to vary across decision makers, both models have in common that they assume a specific form of interaction. In the GHMNL model,  an interaction between the variables $\xb_{ij}$ and $\wb_i$ is present because the linear term takes the form $\xb_{ij}^T\deltab e^{\wb_i^T\gammab}$ (see Equation \ref{eq:V_GHMNL}). In the MXL model, the interaction is included as the linear effect $\zb_{ijt}^T \alphab_i$ contains the term $\zb_{ijt}^T \Delta \wb_i$. In both cases, the interaction can be seen as an interaction generated by effect modification. The effect of $\xb_{ij}$ (or $\zb_{ijt}$) is modified by $\wb_i$, the latter variable is a so-called effect modifier. 

Both models can be embedded into the general framework of varying-coefficient models \citep[see, e.g.,][]{Fan.1999,Hastie.1993,Park.2015}. Although the connection between the MXL model and varying-coefficient models seems not to have been used before, the varying-coefficients framework helps to see that identifiability problems arise if the variables $\zb_{ijt}$ and $\wb_i$ are not distinct. Guided by theoretical expectations about heterogeneity, the researcher applying the MXL model might consider different variables in $\zb_{ijt}$ and $\wb_i$. However, if the underlying theory does not provide deriving such expectations, one faces the challenge of determining which explanatory variables are effect modifiers and which ones represent main effects.  By contrast, the inclusion of the same set of variables in the location and the heterogeneity part of the model does not cause any difficulties in the proposed GHMNL model.

In sum, the benefits of the GHMNL model as compared to the MXL model are:
\begin{itemize}
\item Whereas the MXL model can account for a rather general and unspecific form of heterogeneity without further motivation, the heterogeneity term in the GHMNL model can uncover specific behavioral tendencies. It provides an indicator of the degree of distinctness of choice and measures the strength of heterogeneity in choice behavior.
\item The GHMNL is much sparser in terms of the number of parameters involved and therefore avoids that estimates render unstable without careful variable selection. 
\item It allows deriving a closed-form of the log-likelihood without the need to use simulation methods to obtain choice probabilities, which makes the GHMNL model computationally straightforward.
\item The researcher does not need to decide on a specific and appropriate distribution for the random parameters to approximate the underlying behavioral process. 
\item The GHMNL model avoids identifiability problems and works without repeated measurements. 
\end{itemize}


\tocless
\section{Application:  Spatial Voting and Heterogeneous Electorates}
\noindent
The empirical application uses survey data on electoral choices in multiparty elections to study heterogeneity in spatial voting behavior \citep{Davis.1970, Downs.1957,  Enelow.1984}. Numerous studies have demonstrated that voters evaluate where parties or candidates stand on controversial issues when casting their ballots \citep[see recently, e.g.,][]{Ansolabehere.2018, Jessee.2010, Mauerer.2015}.  The expectation that not all voters behave in the same way but instead differ in their reliance on issues also has a long tradition in the voting literature.  One example is the classic article on voter heterogeneity by \cite{Rivers.1988}, stating that different subgroups of voters apply different choice criteria when voting. Another one is the issue public hypothesis by \cite{Converse.1964}, postulating that the population can be divided into issue publics, each consisting of voters who intensively care about particular issues.

\vspace{1cm}
\subsubsection*{Data}
We draw on the 2017 German parliamentary election study \citep{Rossteutscher.2018} and analyze heterogeneity in voter choice for one of the six major German parties in 2017: the Christian-Democratic Parties (CDU/CSU), the Social-Democratic Party (SPD), the Liberal Party (FDP), the Greens, the Left, and the Alternative for Germany (AfD). Section \ref{app:B} in the Supporting Information contains a detailed description of the measurement and coding of all variables considered in the empirical application.

\subsubsection*{Operationalization of Spatial Proximities}
\noindent
In the tradition of spatial voting approaches, our voter choice model follows the classical proximity model, where the main source of voter utility is the ideological proximity to the parties. Based on a simple linear voter-party proximity specification, we expect that voter $i$ casts a ballot for the party $j$ that offers policy platforms closest to the voter's most preferred positions on $K$ different policy issues. The 2017 German national election study contains three policy issues (immigration, taxes, climate change)  on which the respondents positioned themselves and the parties on eleven-point scales.  Using voter-specific self-placements and perceptions of party placements, the choice-specific variables $z_{ijk}$ in Equation (\ref{eq:V_GHMNL}) contain the absolute proximity between each voter $i$ and party $j$ on each policy issue $k$.    

The empirical application proceeds as follows. Based on previous research on heterogeneity in spatial voting, the first part examines three sources of heterogeneity: issue importance, platform divergence,  and political sophistication.  In the second part, we present the results of a fully-specified voter choice that also accounts for nonpolicy considerations in the voting calculus. 

\tocless
\subsection{Sources of Heterogeneity in Spatial Voting}
\noindent
It has become accepted wisdom that not all voters follow spatial considerations in the same way in making electoral decisions. The debate about heterogeneous electorates has a long tradition in the spatial voting literature. The homogeneity assumption, implying that voters with identical observed characteristics and issue preferences care equally about issues,  has already been questioned in early studies of electoral behavior \citep[see, e.g.,][]{Campbell.1960, Luskin.1987, Meier.1979, Popkin.1991, RePass.1971}. Several concepts, conditions, or sources of heterogeneity have been proposed as to why we should expect systematic individual-level differences in the impact of issue considerations on voting. We empirically examine three theoretical sources of heterogeneity in spatial voting.

\vspace{1cm}
\subsubsection*{Issue Importance}
The first one is the concept of issue importance.  It is the most frequently discussed source of heterogeneity in spatial voting  \citep[see, e.g.,][]{EdwardsIII.1995, Epstein.2000,  Gomez.2001,  Rabinowitz.1982}. If issues are considered as individually salient to voters, then voters are expected to assign these issues a greater weight in the voting-decision process. We employ a typical measure to assess whether  the concept of issue importance provides an explanation to why voters differ in their reliance on issues when voting: the self-reported importance of the three policy issues on five-point scales.

\subsubsection*{Platform Divergence}
Another central condition that must be met so that issues determine voter choice is substantial divergence in offered party positions. Accordingly, voters who see clear differences between parties' policy proposals are expected to rely more strongly on issue attitudes when casting their ballots than those perceiving similar party stands \citep[e.g.,][]{Alvarez.2004, Wessels.2008}. To examine whether platform divergence on an issue causes heterogeneity in the impact of issue considerations on party choice, we employ a subjective measure.  We use the individually perceived range of party positions to identify the degree of platform divergence. The measure is constructed as follows:  For each voter and issue, we first identified the two parties that are perceived to take the most extreme positions on both ends of the issue scales. Then, we computed the absolute difference between these party positions. This results in eleven-point scales, where 0 indicates minimum platform divergence (i.e., all parties are perceived to offer the same position) and 10 maximum platform divergence (i.e., voters perceive the party positions be spread across the entire original eleven-point issue scale).  

We specify a separate model for each of the three policy issues to examine whether voters exhibit heterogeneous reactions to issues due to issue importance and platform convergence. In each model, the location term in Equation (\ref{eq:V_GHMNL}) contains the party-specific constants and spatial proximity.  To identify the constants, we use the CDU as the reference party.  In the heterogeneity term, we consider the concepts of issue importance and platform divergence. Since the heterogeneity term affects the complete location term,  and both sources of heterogeneity in spatial voting are specific to each policy issue, the issue-by-issue model specification allows us to assess whether varying levels of issue importance and platform divergence cause heterogeneous effects.

\begin{table}
\centering
\begin{threeparttable}
\small
\caption{GHMNL Model Estimates, Issue Importance and Platform Divergence} 
\label{tab:1}
\begin{tabular}{@{}p{8,5cm}rrr@{}}
  \toprule
 \textbf{Immigration Issue} & coef. & s.e. & t-value \\ 
  \toprule
\textit{Location Term} &  &  &  \\ 
 SPD & -0.636 & 0.240 & -2.647 \\ 
 FDP & -1.619 & 0.520 & -3.115 \\ 
Greens & -1.081 & 0.382 & -2.829 \\ 
Left & -1.213 & 0.427 & -2.843 \\ 
AfD & -1.361 & 0.439 & -3.098 \\ 
  Issue Proximity & 0.405 & 0.131 & 3.084 \\ 
	\midrule
 \textit{Heterogeneity Term }&  &  &  \\ 
  Issue Importance & 0.035 & 0.062 & 0.568 \\ 
 Platform Divergence & -0.066 & 0.027 & -2.392 \\ 
	\midrule
 Log-likelihood & \multicolumn{3}{c} {1443.414}\\ 
	\bottomrule 
	&&&\\
	\toprule
 \textbf{Tax Issue} & coef. & s.e. & t-value \\ 
  \toprule
	\textit{Location Term} &  &  &  \\ 
SPD & -0.347 & 0.155 & -2.236 \\ 
FDP & -0.784 & 0.314 & -2.497 \\ 
Greens & -0.697 & 0.272 & -2.558 \\ 
Left & -0.673 & 0.285 & -2.361 \\ 
AfD & -0.552 & 0.234 & -2.365 \\ 
  Issue Proximity  & 0.263 & 0.104 & 2.533 \\ 
 	\midrule
 \textit{Heterogeneity Term }&  &  &  \\ 
  Issue Importance & 0.179 & 0.093 & 1.933 \\ 
Platform Divergence  & -0.096 & 0.026 & -3.727 \\ 
 \midrule
 Log-likelihood & \multicolumn{3}{c} {1467.666}  \\ 
\bottomrule 
	&&&\\
	\toprule
 \textbf{Climate Change Issue} & coef. & s.e. & t-value \\ 
\toprule
\textit{Location Term} &  &  &  \\ 
SPD & -0.573 & 0.191 & -3.000 \\ 
 FDP& -0.965 & 0.309 & -3.119 \\ 
Greens& -1.032 & 0.331 & -3.122 \\ 
Left& -1.120 & 0.356 & -3.145 \\ 
AfD & -0.685 & 0.226 & -3.032 \\ 
  Issue Proximity & 0.338 & 0.111 & 3.055 \\ 
   	\midrule
 \textit{Heterogeneity Term }&  &  &  \\ 
  Issue Importance  & 0.112 & 0.067 & 1.669 \\ 
Platform Divergence  & -0.065 & 0.024 & -2.720 \\ 
  \midrule
 Log-likelihood & \multicolumn{3}{c} {1432.210} \\ 
 \bottomrule
\end{tabular}
\begin{tablenotes}
      \footnotesize
       \item \textit{Source:} 2017 German election study \citep{Rossteutscher.2018}. N = 910.
			\item \textit{Note:} Dependent variable is voter choice.   CDU is used as reference party to identify the constants in the location term. 
    \end{tablenotes}
  \end{threeparttable}
\end{table}
Table \ref{tab:1} reports the results. The first column gives the log odds, followed by standard errors and t-values.  The parameters related to the issue proximities in the location term all take positive values and are statistically different from zero at the 5\% significance level. In line with spatial voting approaches, the estimates indicate that the closer voters perceive the parties to their ideal points on the issues, the higher the weight they assign to them when voting, ceteris paribus. The issues of immigration and climate change exhibit the most substantial impact on party choice.

Inspecting the estimates on issue importance and platform divergence in the heterogeneity term reveals interesting choice behavior. In all three models, the coefficients related to the concept of issue importance are positive. Whereas the parameter in the immigration-issue model does not reach conventional statistical significance levels, the parameters in the remaining models do (10\% significance level).  The positive estimates suggest that those voters who consider the tax or the climate change issue individually salient have more distinct party choice preferences, ceteris paribus. In line with previous research \citep[e.g.,][]{EdwardsIII.1995, Rabinowitz.1982},  our model estimates indicate that voters for whom the issues are personally important distinctly prefer specific parties and assign the issues a greater weight in the voting-decision process. By contrast, all coefficients related to the concept of platform divergence take negative values. The negative parameters,  which are all statistically different from zero at the 5\% significance level, indicate heterogeneity in choice behavior. In accord with previous studies \citep[e.g.,][]{Alvarez.2004, Wessels.2008}, the estimates imply that voters who perceive substantial divergence in party positions are more heterogeneous in choosing one party.

\subsubsection*{Political Sophistication}
A large research body has also argued that heterogeneity in issue voting is the result of differences in political sophistication or awareness \citep[see, e.g.,][]{Carmines.1980, Carpini.1993, Gerber.2015, Luskin.1987, Macdonald.1995, Palfrey.1987}. To identify voter segments that might be more sensitive toward issues due to political sophistication, we consider three typical operationalizations of this concept: the stated strength of political interest, objective political knowledge, and education. The level of political interest is measured by relying on voters' self-reports on a five-point scale. Political knowledge is measured using factual knowledge questions with right or wrong answers. Based on the respondents' replies to seven questions, we generated an additive index. We assigned a value of one for each correct answer; wrong and ``don't  know" responses give a value of zero.  Education is a binary variable that takes the value of 1 when the respondent has a higher education entrance qualification and 0 otherwise.

\begin{table}
\centering
\begin{threeparttable}
\small
\caption{GHMNL Model Estimates, Political Sophistication} 
\label{tab:2}
\begin{tabular}{@{}p{8,5cm}rrr@{}}
  \toprule
 \textbf{Immigration Issue} & coef. & s.e. & t-value \\ 
  \toprule
\textit{Location Term} &  &  &  \\ 
SPD & -0.587 & 0.199 & -2.948 \\ 
FDP & -1.592 & 0.437 & -3.644 \\ 
Greens & -1.123 & 0.342 & -3.279 \\ 
Left & -1.190 & 0.359 & -3.314 \\ 
AfD & -1.359 & 0.396 & -3.428 \\ 
Issue Proximity & 0.375 & 0.094 & 4.006 \\ 
\midrule
\textit{Heterogeneity Term }&  &  &  \\  
Political Interest & 0.035 & 0.078 & 0.444 \\ 
Political Knowledge & -0.081 & 0.040 & -2.015 \\ 
Education & -0.082 & 0.136 & -0.605 \\ 
\midrule
 Log-likelihood & \multicolumn{3}{c} {1443.857} \\ 
\bottomrule 
	&&&\\
\toprule
 \textbf{Tax Issue} & coef. & s.e. & t-value \\ 
  \toprule
	\textit{Location Term} &  &  &  \\ 
SPD & -0.372 & 0.141 & -2.641 \\ 
FDP& -0.866 & 0.299 & -2.898 \\ 
 Greens & -0.797 & 0.281 & -2.838 \\ 
 Left& -0.764 & 0.274 & -2.788 \\ 
AfD & -0.664 & 0.239 & -2.782 \\ 
Issue Proximity & 0.263 & 0.077 & 3.432 \\ 
\midrule
\textit{Heterogeneity Term }&  &  &  \\ 
  Political Interest & -0.025 & 0.078 & -0.326 \\ 
  Political Knowledge & 0.049 & 0.050 & 0.981 \\ 
  Education & -0.242 & 0.147 & -1.647 \\ 
	 \midrule
 Log-likelihood & \multicolumn{3}{c} {1473.266}  \\ 
  \bottomrule 
	&&&\\
	\toprule
 \textbf{Climate Change Issue} & coef. & s.e. & t-value \\ 
\toprule
\textit{Location Term} &  &  &  \\ 
SPD& -0.604 & 0.171 & -3.531 \\ 
 FDP& -1.008 & 0.291 & -3.464 \\ 
Greens & -1.092 & 0.297 & -3.681 \\ 
Left & -1.162 & 0.319 & -3.648 \\ 
AfD& -0.711 & 0.223 & -3.188 \\ 
  Issue Proximity & 0.335 & 0.085 & 3.957 \\ 
  \midrule
 \textit{Heterogeneity Term }&  &  &  \\ 
  Political Interest& 0.021 & 0.073 & 0.293 \\ 
  Political Knowledge & -0.010 & 0.039 & -0.251 \\ 
  Education & -0.007 & 0.123 & -0.056 \\ 
	  \midrule
 Log-likelihood & \multicolumn{3}{c} {1437.058}  \\ 
 \bottomrule
\end{tabular}
\begin{tablenotes}
      \footnotesize
      \item \textit{Source:} 2017 German election study \citep{Rossteutscher.2018}. N = 910.
			\item \textit{Note:} Dependent variable is voter choice.   CDU is used as reference party to identify the constants in the location term.  
    \end{tablenotes}
  \end{threeparttable}
\end{table}
Table \ref{tab:2} reports the estimation results. Issue by issue, we specify a model that includes the three measures of political sophistication in the heterogeneity term. The location term again contains constants and spatial proximity. For the immigration-issue model, the coefficient related to political knowledge is negative and statistically different from zero at the 5\% significance level. This result indicates that those voters who have a higher level of political knowledge tend to react more heterogeneously to the immigration issue. Whereas none of the political sophistication measures explain heterogeneous reactions on the climate change issue, the parameter related to education in the tax-issue model is different from zero at the 5\% significance level. Again, the parameter is negative, suggesting that voters with higher education show heterogeneity in voter choice. 

\tocless\subsection{Fully-Specified Voter Choice Model}
\noindent
Next, we present the results of a fully-specified voter choice model. In addition to spatial proximities, the model also contains chooser-specific variables $s_{im}$  in the location term. These are socioeconomic voter characteristics. They account for the importance of voter's nonpolicy motivations in the voting calculus, which presents a central extension of the spatial voting model \citep[see, e.g.,][]{Adams.2005, Merrill.2001}. As nonpolicy factors $\sb_{i}$, we consider four dummy-coded voter attributes in the location term:  worker, religious denomination,  gender, and a regional variable, indicating whether the respondent resides in former West or East Germany.  In the heterogeneity term, we include gender and the regional variable to examine whether there are systematic gender or regional differences in choice behavior.  

Again, we use the CDU as the reference party. The voter choice model is based on 30 degrees of freedom: $3$ issue proximities, $6-1$ constants, and $(6-1) \times 4$ parameters related to voter attributes in the location term, and 2 coefficients in the heterogeneity term. The maximum likelihood point estimates and associated standard errors for the issue proximities  (immigration, tax, climate change) are as follows: $\hat{\alphab}^T= (\hat{\alpha}_{1}, \, \hat{\alpha}_{2}, \, \hat{\alpha}_{3} ) = ( 0.196, 0.202, 0.245)$;  $\hat{\boldsymbol{\Sigmab^{1/2}}}^T_{\alphab}= (\hat{\sigma}_{1}, \, \hat{\sigma}_{2}, \, \hat{\sigma}_{3}) = (0.026, 0.031, 0.034)$.  

Table \ref{tab:3} reports the estimates for the voter attributes in the location and heterogeneity term. In the location term, the interpretation of the coefficients refers to the CDU as this party is used as the reference alternative to identify the model. For example, in line with central social cleavage structures in Germany, Catholics tend to prefer the Christian-Democratic Party CDU compared to the left parties SPD and the Left, ceteris paribus.

\begin{table}
\centering
\begin{threeparttable}
\small
\caption{GHMNL Model Estimates, Fully-specified Voter Choice Model} 
\label{tab:3}
\begin{tabular}{@{}p{4cm}rrrrrr@{}}
  \toprule
& \multicolumn{5}{c} {\textbf{Location }} & \multicolumn{1}{c} {\textbf{Heterogeneity}}  \\
& \multicolumn{5}{c} {\textbf{Term}}& \multicolumn{1}{c} {\textbf{Term}} \\ 
  \toprule
 & SPD & FDP & Greens & Left & AfD &  \\ 
	\cmidrule{2-6}
\textit{Constants} &  &  &  &  &  &  \\ 
  coef. & -0.487 & -0.589 & -1.253 & -0.325 & -0.064 &  \\ 
  s.e. & 0.219 & 0.254 & 0.293 & 0.237 & 0.232 &  \\ 
  t-value & -2.223 & -2.324 & -4.273 & -1.371 & -0.276 &  \\ 
	\cmidrule{1-6} 
 \multicolumn{7}{l}{\hspace{-0.3cm} \textit{Worker} (1 worker, 0 otherwise)}  \\ 
  coef. & 0.095 & -1.154 & -0.494 & -0.154 & -0.177 &  \\ 
  s.e.& 0.262 & 0.460 & 0.382 & 0.329 & 0.321 &  \\ 
  t-value & 0.363 & -2.506 & -1.291 & -0.468 & -0.552 &  \\ 
	\cmidrule{1-6}
\multicolumn{7}{l}{\hspace{-0.3cm} \textit{Religious Denomination} (1 Catholic, 0 otherwise)} \\
  coef. & -0.748 & -0.287 & -0.260 & -0.685 & -0.590 &  \\ 
  s.e. & 0.236 & 0.247 & 0.235 & 0.298 & 0.315 &  \\ 
  t-value  & -3.174 & -1.160 & -1.105 & -2.299 & -1.874 &  \\ 
	\cmidrule{1-7}
\multicolumn{7}{l}{\hspace{-0.3cm} \textit{Gender} (1 female, 0 male)}   \\ 
  coef. & -0.206 & -0.734 & -0.200 & -0.265 & -0.533 & -0.239 \\ 
  s.e. & 0.219 & 0.329 & 0.256 & 0.279 & 0.309 & 0.120 \\ 
  t-value  & -0.941 & -2.233 & -0.779 & -0.950 & -1.723 & -1.997 \\ 
	\cmidrule{1-7}
\multicolumn{7}{l}{\hspace{-0.3cm} \textit{Region} (1 former West Germany, 0 former East Germany)}   \\ 
  coef. & 0.273 & 0.022 & 0.616 & -0.512 & -0.533 & 0.233 \\ 
  s.e.& 0.234 & 0.297 & 0.313 & 0.285 & 0.285 & 0.122 \\ 
   t-value  & 1.166 & 0.074 & 1.966 & -1.794 & -1.869 & 1.914 \\ 
 \bottomrule
\end{tabular}
\begin{tablenotes}
      \footnotesize
      \item \textit{Source:} 2017 German election study \citep{Rossteutscher.2018}. N = 910. LogL = 1289.378.
			\item \textit{Note:} Dependent variable is voter choice.   CDU is reference party.  The estimates for the spatial proximities are reported in the text. 
    \end{tablenotes}
  \end{threeparttable}
\end{table}
Regarding the heterogeneity term, the coefficients are not specific to a particular party. The corresponding effects are global and do not relate to a reference alternative. The coefficient associated with the gender variable is negative and statistically different from zero at the 5\% significance level.  The negative value indicates that females show more heterogeneity in voter choice than males, ceteris paribus. By contrast, the coefficient related to the regional variable is positive and statistically different from zero at the 10\% significance level. This result suggests that voters living in former West Germany have more distinct party choice preferences than those residing in East Germany, ceteris paribus.

\vspace{1cm}
\tocless
\section{Conclusion}
\noindent
Categorical dependent variables are widespread in political science, and the discipline has contributed enormously to methods for the analysis of nominal responses.  Political scientists studying nominal-scaled dependent variables as a choice among discrete alternatives frequently hypothesize heterogeneous effects. Being guided by recommendations from the political methodology literature, current practice to study heterogeneity in choice behavior is to allow the parameters associated with choice-specific attributes to vary randomly across decision makers. In particular, political science settled on the mixed logit model. As we have demonstrated,  the mixed logit model, however, comes with several drawbacks, such as a high number of parameters to be estimated, identifiability problems, or the need to specify a specific and appropriate distribution for the random effects.

Building on the standard MNL model,   a general multinomial logit model for the systematic study of heterogeneity in choice behavior is proposed, which avoids these difficulties, is computationally straightforward, and comes with convenient properties.  The proposed GHMNL model integrates a heterogeneity term into the systematic part of the utility function and accounts for behavioral choice tendencies without referring to latent variables.  The heterogeneity term is linked to explanatory variables, indicates the degree of distinctiveness of choice or the impact of heterogeneity in choice behavior.

Drawing on theoretical sources of heterogeneity in spatial voting (issue importance, platform divergence, and political sophistication), we have demonstrated how the GHMNL model allows incorporating theoretical expectations into the empirical modeling and how it can improve our understanding of heterogeneous electorates, which remains to be an important topic in electoral research \citep[e.g.,][]{Basinger.2005, Federico.2013, Gerber.2015,  Peterson.2005, Singh.2014b}.  For example, our empirical estimates suggest that voters who consider the issues are personally important distinctly prefer specific parties and assign the issues a greater weight in the voting-decision process. By contrast, platform divergence induces heterogeneity in spatial voting behavior. Depending on the measure and the issue under consideration,  our results also indicate that the higher the level of political sophistication, the more voters tend to exhibit heterogeneous reactions to issues. As many research questions in political science involve theoretical expectations about heterogeneity, we see a wide range of applications in all political science sub-disciplines.

\vspace{1cm}
\bibliographystyle{AJPS_Citation2}
\bibliography{../../Literatur/Bib_Files/bib_6}

\tocless
\blanco{
\section{Supporting Information}
\noindent
Additional Supporting Information can be found in the Online Appendix: 

\begin{itemize}
	\item[] Appendix A: R Function
	\item[]	Appendix B: Empirical Application: Data and Operationalization
\end{itemize}
\end{center}
}
\newpage
\appendix

\linespread{1.0}\selectfont
\setcounter{table}{0}
\renewcommand{\thetable}{A\arabic{table}}
\setcounter{page}{1}

\blanco{
\begin{center}
{\Large \textbf{Supporting Information}}\\
on \\
\textit{\Large{Heterogeneity in General Multinomial Choice Models}}
}
\clearpage

\vspace{1cm}


\noindent

\section{R Function}\label{app:A}
The R function to fit the \textit{General Heterogeneous Multinomial Logit Model} (GHMNL) is 

\textit{GHMNL(dat, namesglob, Indglob, namescats, Indcats, nameshet, nameresp, k, pen)}.

The elements of the function are: 
\begin{itemize}
\item[] dat:  data set in long format
\item[] namesglob: names of global variables (chooser-specific attributes)
\item[] Indglob $>$0: global variables are included, otherwise ignored
\item[] namescats:  names  of category-specific  variables (choice-specific attributes)
\item[] Indcats $>$0: category-specific  variables are included, otherwise ignored
\item[] nameshet: names of variables in the heterogeneity term (can be empty)
\item[] namesresp: name of response variable (vector of 0-1 dummy variables, based on the nominal dependent variable)
\item[] k: number of alternatives
\item[] pen: unpenalized estimates are obtained by choosing pen=0; if pen $> 0$ a ridge penalized estimate is computed; should always be small, for example, pen = 0.0000001 
\end{itemize}

\section{Empirical Application: Data and Operationalization}\label{app:B}
Our empirical application uses the 2017 German national election study \citep{Rossteutscher.2018}. We restrict the analysis to voter choice for one of the six major German parties  using the second vote: the Christian-Democratic Parties (CDU/CSU)\footnote{For voters in Bavaria the CSU instead of the CDU is used.}, the Social-Democratic Party (SPD), the Liberal Party (FDP), the Greens, the Left Party, and the Alternative for Germany (AfD). 

\subsection{Spatial Proximities}
The election study contains three policy issues (immigration, taxes, climate change) on which the respondents positioned themselves and the parties on eleven-point scales. The end points are labeled as follows: 
\begin{itemize}
	\item Immigration: 1 = ``immigration should be facilitated", 11 = ``immigration should be restricted"; 
	\item Taxes: 1 =  ``lower taxes, even if this means a reduction in the benefits offered by the social state", 11 = ``more benefits offered by the social state, even if this means an increase in taxation"; 
	\item Climate Change: 1 =  ``fight against climate change should take precedence, even if it impairs economic growth", 11 = ``economic growth should take precedence, even if it impairs the fight against climate change".  
\end{itemize}
Based on this information, we constructed  the choice-specific variables  $\zb_{ij}$.  For each issue $k$, the voter-party proximity measures are defined as the absolute negative distance between respondent-specific perceptions of party positions ($p_{ijk}$) and self-reported respondent position $(x_{ik})$ so that $z_{ijk} \defas -|x_{ik} - p_{ijk}|$. 

\subsection{Voter Attributes to Account for Nonpolicy Considerations}
As nonpolicy factors $\sb_{i}$, we consider voter attributes that reflect central social cleavage structures influencing party choice in Germany: 
\begin{itemize}
	\item worker: 1 (worker), 0 (otherwise); 
	\item religious denomination: 1 (catholic), 0 (otherwise); 
	\item gender: 1 (female), 0 (male); 
	\item West/East Germany: 1 (former West Germany), 0 (former East Germany).
\end{itemize}

\subsection{Theoretical Sources of  Heterogeneity in Spatial Voting}
The election study also includes sociodemographic information and several questions frequently used to explain variability in issue voting. To identify voter segments which might be more sensitive toward issues, we selected three potential systematic sources of heterogeneity.

\subsubsection*{Issue Importance}
We measure issue importance by relying on self-reports to the survey question ``How important is the [issue]  to you personally?". The respondents were asked to state the importance of the three policy issues on a five-point scale running from ``not at all important" to ``very important". 

\subsubsection*{Platform Divergence}
We use the individually perceived range of party positions to identify the degree of platform divergence. For each voter and issue, we computed the absolute difference between the perceived most extreme position on one and the other end of the issue scales. This results in eleven-point scales, with 0 indicating minimum platform divergence and 10 maximum platform divergence.  

\subsubsection*{Political Sophistication}
We measure the concept of political sophistication by three standard measures: the stated strength of political interest, objective political knowledge, and education. The level of political interest is measured by relying on voters' self-reports on a five-point scale. Political knowledge is measured using factual knowledge questions with right or wrong answers. Based on the respondents' replies to seven questions, we generated an additive index in which for each correct answer, a value of one is assigned, whereas wrong and ``don't  know" answers give a value of zero. The index is based on two questions about the German electoral system (survey questions: ``Which one of the two votes is decisive for the relative strengths of the parties in the Bundestag?"; ``What is the percentage of the second vote a party needs to be able to send delegates to the Bundestag definitely?") and two questions regarding the budget deficit and the unemployment rate. In addition, the respondents were confronted with pictures showing three politicians and were asked to state the party each politician belongs to. These politicians are Martin Schulz (SPD), Katrin Goering-Eckardt (Greens), and Christian Lindner (FDP). The answers are aggregated by counting the correct responses, yielding an eight-categorical variable (0 none correct, 7 all correct). Education is a dichotomous variable that takes the value of 1 when the respondent has a higher education entrance qualification (i.e., a higher-school certificate with university admission) and 0 otherwise.

\end{document}